\documentclass[
reprint,
twocolumn,
amsmath,amssymb,
aps,
prx,
floatfix,
longbibliography
]{revtex4-2}

\usepackage{graphicx}
\usepackage{dcolumn}
\usepackage{bm}
\usepackage{braket}
\usepackage{siunitx}

\def\statefidelity{\qty{98.2}{\%}}
\def\EPC{\num{3.25\pm0.09d-4}}
\def\ef{\num{3d-3}}
\def\readoutfidelityGMM{\qty{85.8}{\%}}
\def\readoutfidelityDNN{\qty{93.8}{\%}}

\begin{document}

\title{Systematic study of High $\mathbf{E_J/E_C}$ transmon qudits up to $\mathbf{d=12}$}
\author{Zihao Wang}
\author{Rayleigh W. Parker}
\author{Elizabeth Champion}
\author{Machiel S. Blok}
\email[]{machielblok@rochester.edu}
\affiliation{Department of Physics and Astronomy, University of Rochester, Rochester, NY 14627}
\date{\today}

\begin{abstract}
Qudits provide a resource-efficient alternative to qubits for quantum information processing. The multilevel nature of the transmon, with its individually resolvable transition frequencies, makes it an attractive platform for superconducting circuit-based qudits. In this work, we systematically analyze the trade-offs associated with encoding high-dimensional quantum information in fixed-frequency transmons. Designing high $E_J/E_C$ ratios of up to 325, we observe up to 12 levels ($d=12$) on a single transmon. Despite the decreased anharmonicity, we demonstrate process infidelities $e_f < \ef$ for qubit-like operations in each adjacent-level qubit subspace in the lowest 10 levels. Furthermore, we achieve a 10-state readout assignment fidelity of \readoutfidelityDNN\ with the assistance of deep neural network classification of a multi-tone dispersive measurement. We find that the Hahn echo time $T_{2E}$ for the higher levels is close to the limit of $T_1$ decay, primarily limited by bosonic enhancement. We verify the recently introduced Josephson harmonics model, finding that it yields better predictions for the transition frequencies and charge dispersion. Finally, we show strong $ZZ$-like coupling between the higher energy levels in a two-transmon system. Our high-fidelity control and readout methods, in combination with our comprehensive characterization of the transmon model, suggest that the high-$E_J/E_C$ transmon is a powerful tool for exploring excited states in circuit quantum electrodynamics.
\end{abstract}

\maketitle

\section{\label{sec:intro}Introduction}

Superconducting circuits are a leading platform for quantum information processing \cite{blaisCircuitQuantumElectrodynamics2021, krantzQuantumEngineerGuide2019}. Among the many varieties of superconducting circuits, transmons \cite{koch2007ChargeinsensitiveQubitDesign} are popular for their simple structure and potential for scalability. Recent works have made remarkable progress toward fault-tolerant quantum computing using transmons \cite{kimEvidenceUtilityQuantum2023, acharya2023SuppressingQuantumErrors}. In most experiments transmons are operated as qubits: the computational subspace is spanned by the two lowest-energy eigenstates $\ket{0}$ and $\ket{1}$, and all other energy levels are expected to be unoccupied. In reality, however, transmons are weakly anharmonic systems with multiple eigenstates. The effects of the excited states of the transmon can be detrimental to the computing system unless mitigated \cite{miao2023}. 

One can instead choose to use the transmon's higher energy levels as computational states, thereby operating it as a qudit \cite{bianchettiControlTomographyThree2010, petererCoherenceDecayHigher2015, lescanne2019EscapeDrivenQuantum, kristen2020, wu2020, chenTransmonQubitReadout2023, nguyenEmpoweringHighdimensionalQuantum2023, seifertExploringQuquartComputation2023a, kehrerImprovingTransmonQudit2024}. Qudits are advantageous in part because they encode a Hilbert space with dimension $d > 2$ in a device with the same physical footprint as a qubit. Qudit-based quantum algorithms \cite{kiktenko2023} can be used to simplify the decomposition of multi-qubit gates and reduce the circuit depth. In addition to superconducting circuits, high-fidelity control and readout of qudit systems has been achieved on platforms such as trapped ions \cite{ringbauerUniversalQuditQuantum2022, hrmo2023}, NV centers \cite{fu2022}, molecular spins \cite{chiesa2023}, and silicon-photonics \cite{chi2022}. 

Transmon qudits are an active area of research, with experiments including demonstrations of high-fidelity control and tomography \cite{bianchettiControlTomographyThree2010, wu2020}, the observation of escape into unconfined states \cite{lescanne2019EscapeDrivenQuantum}, quantum information scrambling \cite{blokQuantumInformationScrambling2021}, the encoding of multiple qubits \cite{cao2024}, and the emulation of a large quantum spin \cite{champion2024, neeleyEmulationQuantumSpin2009}. Advanced qudit gates have also been proposed \cite{wu2020, fischer2023, luo2023, nguyenEmpoweringHighdimensionalQuantum2023} and used for prototypical quantum algorithms \cite{liu2023}. However, most transmon qudit experiments were limited to $d=4$ due to the trade-off in the transmon spectrum between large qudit dimension versus coherence, anharmonicity, and readout fidelity. These factors are governed by the Josephson energy $E_J$ and charging energy $E_C$, which determine the parameter space for transmon qudits. 

Here, we explore a range of large $E_J/E_C$ fixed frequency transmons and systematically evaluate their qudit performance. We demonstrate high-fidelity control and state preparation on transmons with $E_J/E_C$ ratios up to 325 and Hilbert space dimensions up to $d=12$. Using frequency-multiplexed readout on a single resonator for each transmon, we realize single-shot readout of up to 10 levels with \readoutfidelityDNN\ assignment fidelity. We present an analysis of the coherence of the higher levels that illustrates how the scaling of energy relaxation with excitation number may help to distinguish different noise processes. Finally, we compare our experimental results with the traditional transmon model and find that the correction from the recently proposed Josephson harmonics model \cite{willschObservationJosephsonHarmonics2024} leads to better predictions for transition frequencies and charge dispersion. Our study aims to understand and quantify the various trade-offs in transmon qudit design and demonstrates high-fidelity control and readout of high-dimensional qudits. In the future, large $E_J/E_C$ transmons may be used for qudit-based quantum information processing \cite{kiktenko2023} and quantum simulation of high-dimensional systems \cite{georgescu2014, bauer2023}. They could also serve to validate models that aim to explain e.g. Josephson harmonics or transmon ionization \cite{willschObservationJosephsonHarmonics2024, shillitoDynamicsTransmonIonization2022, dumas2024}. 


\section{\label{sec:concept}High-$E_J/E_C$ transmons}

We begin with a brief review of the standard transmon Hamiltonian \cite{koch2007ChargeinsensitiveQubitDesign},
\begin{equation}
\hat{H}_{\text{st}} = 4 E_C (\hat{n} - n_g)^2 - E_J \cos\hat{\phi},
\label{eq:standard_transmon_hamiltonian}
\end{equation}
where $\hat{n}$ and $\hat{\phi}$ are the reduced charge and phase operators and $n_g$ is the offset charge. The Josephson energy $E_J$ and charging energy $E_C$ are two free parameters set by the size of the Josephson junction and the geometry of the transmon, which can be changed in design and fabrication. In our experiments we use fixed-frequency transmons, composed of a single Josephson junction and a large capacitor. In the transmon regime, where typically $E_J/E_C \geq 50$, the base transition frequency $f_{01}$ and anharmonicity $\alpha \equiv f_{12} - f_{01}$ can be calculated from $E_J$ and $E_C$ through
\begin{equation}
hf_{01} \simeq \sqrt{8 E_J E_C} - E_C, \quad h\alpha \simeq -E_C,
\label{eq:f01_alpha}
\end{equation}
\begin{equation}
\frac{f_{01}}{|\alpha|} \simeq \sqrt{\frac{8E_J}{E_C}}.
\label{eq:alphar}
\end{equation}
Conversely, we can use these relations to estimate a transmon's $E_J$ and $E_C$ from its experimentally measured $f_{01}$ and $\alpha$. 

Charge qubits like transmons require careful consideration of charge dispersion $\epsilon_m$, which is the maximum fluctuation range of each energy level $m$ as $n_g$ varies and is approximately given by
\begin{equation}
\epsilon_m \simeq (-1)^m E_C \frac{2^{4m+5}}{m!} \sqrt{\frac{2}{\pi}} \left(\frac{E_J}{2E_C}\right)^{\frac{m}{2}+\frac{3}{4}} e^{-\sqrt{8E_J/E_c}}.
\label{eq:charge_dispersion}
\end{equation}
For the case where $E_J/E_C < 50$ \cite{koch2007ChargeinsensitiveQubitDesign}, the large $\epsilon_m$ can be a strong dephasing channel, limiting the coherence time. Hence, reducing $\epsilon_m$ usually requires a large value of $E_J/E_C$. However, as shown in Eq. (\ref{eq:alphar}), for a fixed target $f_{01}$, increasing $E_J/E_C$ will also reduce anharmonicity, which may cause leakage and introduce more coherent error for fast drives. The base transition frequency $f_{01}$ of a transmon is typically set between 3 and \qty{6}{\giga \hertz} \cite{seifertExploringQuquartComputation2023a, acharya2023SuppressingQuantumErrors} with $E_J/E_C$ in the range of 50-100. These parameters balance these two sources of error while taking advantage of the availability of microwave electronic components in this frequency range. Increasing the base transition frequency along with $E_J/E_C$ can reduce both dephasing and leakage errors, but may suffer from strong quasiparticle and dielectric losses, and will therefore generally have a short lifetime. This trade-off is still under study \cite{anferov2024}.

\begin{figure}[tp]
\includegraphics{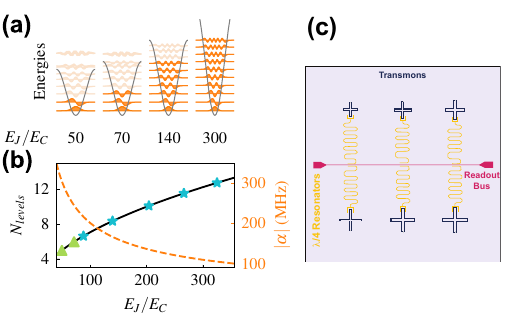}
\caption{
\label{fig:1_concept} 
High-$E_J/E_C$ transmons. (a) The cosine potential wells (grey) and lowest 10 eigenstate wavefunctions (orange) for transmons with $f_{01}=\qty{5}{\GHz}$ and different $E_J/E_C$ ratios. The eigenenergies are represented by the vertical positions of the wavefunctions. (b) The relation between $N_{\text{levels}}$ (black), $|\alpha|$ (orange) and $E_J/E_C$ for a transmon with a fixed base transition frequency of $f_{01}=\qty{5}{\GHz}$. The scatter points correspond to $E_J/E_C$ values typical of qubit and qutrit experiments (green triangles) and those used in this work (cyan stars). (c) One of the chip designs used in this work. Each transmon couples to a readout resonator and is driven and read out through the same transmission line (red). The smaller $E_C$ of the high-$E_J/E_C$ transmon yields a larger footprint compared to the low-$E_J/E_C$ transmon due to its larger capacitance $C$.
}
\end{figure}

In our experiments we set out to observe and utilize as many levels as possible. In Fig. \ref{fig:1_concept}(a) we show the lowest 10 eigenenergies of transmons with $f_{01}=\qty{5}{\GHz}$ and different $E_J/E_C$ ratios. In each case the cosine potential in Eq. (\ref{eq:standard_transmon_hamiltonian}) can only confine a finite number of energy levels within the potential well. The levels close to and above the top of potential well have wavefunctions that are delocalized in phase: these eigenstates are close to the charge eigenstates and are therefore sensitive to charge noise. Only the low-energy levels deep inside the potential well behave like eigenstates of an anharmonic oscillator and are protected from charge noise. In the transmon regime, the number of such confined levels $N_{\text{levels}}$ can be approximated by the ratio between the depth of potential well and the level spacing \cite{shillitoDynamicsTransmonIonization2022}
\begin{equation}
N_{\text{levels}} \simeq \frac{2E_J}{\sqrt{8E_JE_C}} = \sqrt{\frac{E_J}{2E_C}} \simeq \frac{1}{\phi_{zpf}^2},
\label{eq:N_levels}
\end{equation}
where $\phi_{zpf}$ is the zero-point fluctuation of phase operator. As shown in Eq. (\ref{eq:N_levels}), we expect a high-$E_J/E_C$ transmon to have a large $N_{\text{levels}}$ where each level is more localized inside the potential well compared to a low-$E_J/E_C$ transmon. We achieve this on our device by increasing $E_J$ to raise the height of the potential well and decreasing $E_C$ to keep the transition frequencies in a typical frequency range. In Fig. \ref{fig:1_concept}(b), we show a numerical calculation \cite{groszkowski2021} of the relation between $|\alpha|$, $E_J/E_C$, and $N_{\text{levels}}$ for a transmon with fixed $f_{01}=\qty{5}{\GHz}$. The value of $E_J / E_C$ is a crucial transmon design parameter that depends on the intended use. $E_J/E_C \sim 50$ is commonly used for transmon qubits ($d=2$), while a ratio around 70 is typical for qutrit ($d=3$) computation. Our transmons (cyan stars) go well beyond this regime, leading to a reduced sensitivity to charge noise and allowing access to highly excited states at the cost of a reduced anharmonicity, see Fig. \ref{fig:1_concept}(b).

In this paper we use six fixed-frequency transmons ($Q_0-Q_5$) across three different chips, one of which is shown in Fig. \ref{fig:1_concept}(c). We use the Xmon design \cite{barends2013} with $E_J/E_C$ values ranging from 88 to 325. Their base transition frequencies $f_{01}$ are all around \qty{5}{\GHz}, and each transmon is coupled to a readout resonator with a resonant frequency around 6-\qty{7}{\GHz}. The lowest-$E_J/E_C$ transmon ($Q_0$) has an anharmonicity of $\alpha=\qty{-209}{\MHz}$, while the highest-$E_J/E_C$ transmon ($Q_5$) has $\alpha=\qty{-104}{\MHz}$. The detailed parameters of our devices can be found in Appendix \ref{sec:device_params}.

\section{\label{sec:control}High-fidelity control and state preparation}

\begin{figure}[!t]
\includegraphics{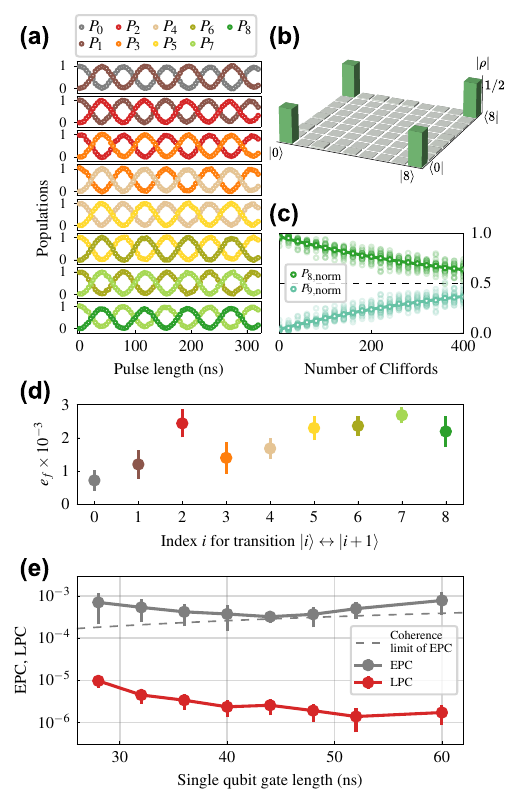}
\caption{
\label{fig:2_control} 
Coherent control of a high-$E_J/E_C$ transmon. (a) Rabi experiments for each transition between adjacent states on $Q_4$. In each case the drive amplitude is calibrated to the same Rabi frequency, corresponding to a \qty{40}{\ns} $\pi$-pulse. (b) Magnitudes of the density matrix elements recovered from qudit state tomography after nominally preparing the superposition state $\ket{0}+\ket{8}$ on $Q_4$ which, yielding a state fidelity of \statefidelity. (c) Example of qubit-like RB on the $\{\ket{8},\ket{9}\}$ subspace of $Q_5$. The populations within the subspace are normalized (circles) prior to being fit (solid line), accounting for decay outside of the subspace. (d) Process infidelities $e_f$ for qubit-like RB on each calibrated transition of $Q_5$. All transitions have $e_f < \ef$. (e) Error per Clifford (EPC, gray dots) and leakage per Clifford (LPC, red dots) recovered from RB on the $\{\ket{0},\ket{1}\}$ subspace of $Q_5$. The detuned-DRAG pulses are recalibrated for each single-qubit gate length. At a pulse length of \qty{44}{\ns} the EPC is \EPC, which is close to coherence limit (gray dashed line).
}
\end{figure}

Every state transition is resolvable in the transmon frequency spectrum. A resonant microwave pulse with frequency $f_{i, i+1}=f_{i+1}-f_{i}$ can drive the transmon between state $\ket{i}$ and $\ket{i+1}$ with eigenenergies $hf_{i}, hf_{i+1}$. In Fig. \ref{fig:2_control}(a) we show Rabi oscillations between neighboring states on $Q_4$ up to the $\ket{7} \leftrightarrow \ket{8}$ transition. While transmon selection rules prohibit directly driving transitions between all states in the qudit, we can realize full control over this large Hilbert space by decomposing any $SU(d)$ unitary into sequential pulses between adjacent energy levels, which form effective qubit subspaces \cite{kononenko2021}.
For each subspace, we use \qty{40}{\ns} flat-top pulses with cosine ramps and a DRAG correction, and the transition frequencies are calibrated using Ramsey experiments. As an example, we prepare the superposition state $(\ket{0}+\ket{8})/\sqrt{2}$ on $Q_4$ and perform state tomography \cite{bianchettiControlTomographyThree2010, qiQuantumStateTomography2013, smolin2012}. We first initialize our transmon to $\ket{0}$ with a heralding measurement. The state preparation begins with an $X^{01}_{\pi/2}$ gate to generate a superposition, followed by $X^{12}_{\pi}, X^{23}_{\pi}, X^{34}_{\pi}, X^{45}_{\pi}, X^{56}_{\pi}, X^{67}_{\pi}, X^{78}_{\pi}$ to reach the target state. We also add a virtual Z rotation as a phase correction between state preparation gates and tomography gates to compensate for spurious phases resulting from the AC Stark shift generated by the state preparation pulses \cite{morvan2021}. In this $d=9$ system, we choose a tomography gate set with 73 different gate sequences and repeat 5000 shots for each gate sequence. We give more details about the state tomography in Appendix \ref{sec:state_tomography}. Our readout method is introduced in Sec. \ref{sec:readout}. The tomography result is shown in Fig. \ref{fig:2_control}(b) where we plot the magnitude of the recovered density matrix for the $(\ket{0}+\ket{8})/\sqrt{2}$ state, yielding a state fidelity of 98.2\%. 
The fidelity here is limited by the accumulated errors of eight physical pulses and the SPAM errors.

To further examine our control quality, we calibrate and perform Randomized Benchmarking (RB) \cite{magesanScalableRobustRandomized2011, mckay2019, jafarzadeh2020, morvan2021, kononenko2021} on the lowest 10 levels of $Q_5$, which has the highest $E_J/E_C$ and weakest anharmonicity. Although qudit RB \cite{jafarzadeh2020} would be the ideal metric to fully characterize errors, it is challenging to implement a long qudit Clifford sequence within the finite coherence time since there are more than 10000 Clifford gates \cite{tolar2018} in the $d=10$ Hilbert space and most of them need to be decomposed into multiple primitive gates in our method. Instead, we perform qubit-like RB \cite{morvan2021} within each two-level subspace spanned by adjacent states $\ket{i} \leftrightarrow \ket{i+1}$ and extract the process infidelity $e_f$, which can effectively quantify the Pauli error in this two level subspace. In Fig. \ref{fig:2_control}(c) we give an example of qubit RB in the $\{\ket{8},\ket{9}\}$ subspace on $Q_5$. In this experiment, we first prepare our transmon in $\ket{8}$, then apply $m+1$ randomly chosen Clifford gates in the $\{\ket{8},\ket{9}\}$ subspace $\mathcal{C}^{89}_0, \mathcal{C}^{89}_1,... , \mathcal{C}^{89}_{m-1}, \mathcal{C}^{89}_m$ where the final gate is chosen to be the inverse of the sequence up to that point. The final result is averaged over 1000 shots and 30 different randomizations for each circuit depth $m$, and we fit the normalized population $P_{8, \text{norm}} = P_8/(P_8+P_9)$ to an exponential decay model $Ar^{m}+C$ before calculating the process infidelity $e_f = (1-r)(1-1/d^2)$ \cite{arute2019}. We run this qubit RB and find $e_f < \ef$ for all calibrated transitions shown in Fig. \ref{fig:2_control}(d). The overall increase in error for higher levels likely arises from their shorter coherence times; see Sec. \ref{sec:coherence}. 

The relatively weak anharmonicities of our high-$E_J/E_C$ transmons poses a trade-off because they may introduce additional leakage errors and stark shifts for fast drives. In a transmon, the frequencies $f_{i,i+1}$ are not equally spaced for each transition. Instead, the $i$-th anharmonicity $\alpha_i \equiv f_{i, i+1} - f_{i-1, i}$ becomes stronger for higher transitions until we reach the top of potential well. For example, $Q_5$ has $\alpha_1 = \qty{-104}{\MHz}$ and $\alpha_9 = \qty{-170}{\MHz}$. To test the impact of anharmonicity on gate fidelity, we run RB experiments with up to 2000 Cliffords in the $\{\ket{0},\ket{1}\}$ subspace for different primitive gate pulse lengths. We use detuned-DRAG pulses, recalibrated for each pulse length, where the DRAG weight is optimized to minimize leakage to $\ket{2}$ and an additional detuning of the pulse frequency is responsible for reducing the phase error in the qubit subspace \cite{chenMeasuringSuppressingQuantum2016}. In Fig. \ref{fig:2_control}(e) we show the error per Clifford (EPC) and leakage per Clifford (LPC) extracted from these RB experiments. We find the minimal $\text{EPC} = \EPC$ for a \qty{44}{\ns} pulse, which is comparable to typical low-$E_J/E_C$ transmon devices with the state of art EPC around $\num{e-4}$ to $\num{e-3}$ \cite{vepsalainen2022, mckay2019, werninghaus2021, kim2023, nguyenEmpoweringHighdimensionalQuantum2023}. This optimal pulse length best balances the contributions from coherent and incoherent errors. A longer pulse would suffer more from decoherence, while a shorter pulse would usually cause more leakage and unwanted AC Stark shifts. We also use independent $T_1$ and $T_2$ measurement results (see Sec. \ref{sec:coherence}) to predict the coherence limit of the EPC, and find that our gate quality approaches the coherence limit for 40-\qty{48}{\ns} pulses.

\section{\label{sec:readout}Multi-tone Readout and state assignment}

\begin{figure*}[tp]
\centering
\includegraphics{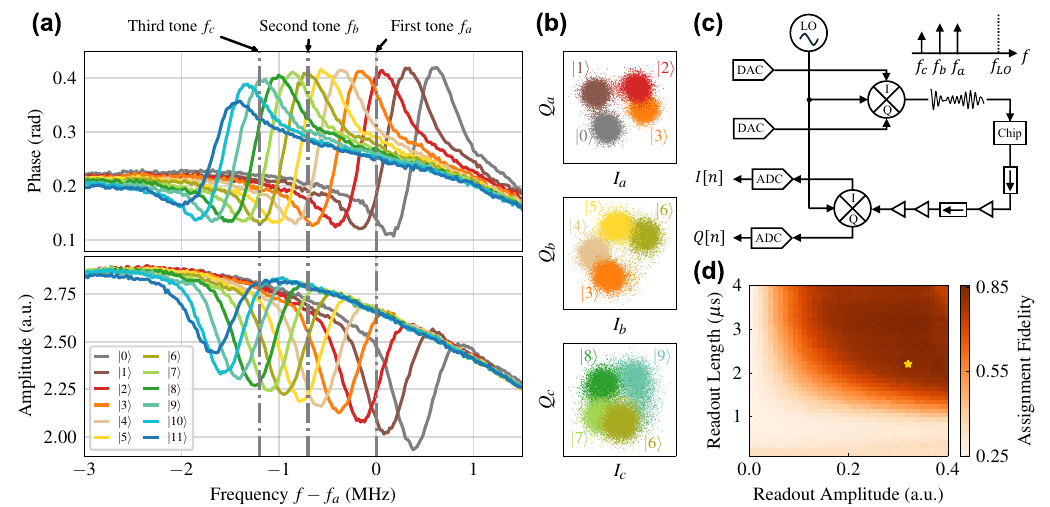}
\caption{
\label{fig:3_RFS} 
Multi-tone single-shot readout of high-$E_J/E_C$ transmon qudits. (a) Spectroscopy of readout resonator $R_5$. The phase (upper) and amplitude (lower) responses are shown when $Q_5$ is prepared in each eigenstate up to $\ket{11}$. The frequencies of the three tones used for multi-tone single-shot readout are indicated by the vertical dotted dashed lines. (b) $IQ$ points in each $I\!-\!Q$ plane, produced by demodulating and integrating the signal at each of the three readout frequencies. We choose these frequencies such that each tone can unambiguously distinguish between four states while the $IQ$ responses for other states overlap. (c) Simplified schematic of the readout setup. The frequency-multiplexed signal is generated at baseband and upconverted to RF, yielding a signal containing $f_a$, $f_b$ and $f_c$ simultaneously. This signal is then sent to the chip, and the signal transmitted past the resonator is amplified by a TWPA, HEMT, and room-temperature amplifier prior to being downconverted to baseband and sampled. Isolators are used to reduce back-propagation noise from amplifier. (d) An example of the pulse amplitude and length calibration for dispersive readout at a single frequency. A fast readout with low photon number (lower left) is ideal, but usually yields a low readout assignment fidelity. We choose a relatively fast readout speed (yellow star) on our device.
}
\end{figure*}

Having characterized our unitary control fidelity, we now turn our attention to readout. One of the long-standing challenges of using qudit systems is the need for high-fidelity single-shot readout of all $d$ levels. Transmon readout is typically performed by capacitively coupling the transmon to a far-detuned linear resonator. The Hamiltonian of such a system is
\begin{equation}
\hat{H}_{t-r} = \hat{H}_{\text{st}} + \hat{H}_r + \hat{V}_{t-r},
\label{eq:standard_transmon_resonator_hamiltonian}
\end{equation}
where
\begin{equation}
\hat{H}_r = h f_r \hat{a}^{\dagger}\hat{a},
\label{eq:resonator_hamiltonian}
\end{equation}
\begin{equation}
\hat{V}_{t-r} = \text{i} hg \hat{n} (\hat{a}^{\dagger} - \hat{a}).
\label{eq:transmon_resonator_interaction}
\end{equation}
In the dispersive limit, where $g \ll |f_{i, i+1} - f_r|$, the Hamiltonian of this system is approximately given by \cite{blaisCircuitQuantumElectrodynamics2021}
\begin{equation}
\hat{H}_U = \hat{H}_r
+ \sum_i h \tilde{f}_i \ket{i}\bra{i} 
+ \sum_i h \chi_i \hat{a}^{\dagger}\hat{a} \ket{i}\bra{i}.
\label{eq:standard_dispersive_hamiltonian}
\end{equation}
Here $\ket{i}$ is the transmon eigenstate, $h \tilde{f}_j$ are the Lamb-shifted transmon energies, and $\chi_i$ is the Stark shift (see Appendix \ref{sec:dispersive} for more details). The last term in Eq. (\ref{eq:standard_dispersive_hamiltonian}) is the effective coupling between the transmon and its readout resonator. Each state $\ket{i}$ of the transmon induces a different frequency shift $\chi_i$ on the resonator such that $f_{r, \ket{i}} = f_r + \chi_i$. By probing the resonator with a near-resonant microwave signal and detecting the change in amplitude and phase of the returned signal, we can infer the transmon state.

The dispersive shift can be seen in Fig. \ref{fig:3_RFS}(a) where we show the spectroscopy result of the resonator $R_5$ for 12 different transmon states of $Q_5$, and we find each of these states induces a distinct resonator frequency response. At each readout frequency, we first prepare the transmon into a certain state $\ket{i}$ before sending a \qty{2.2}{\us} readout pulse to the readout transmission line and detecting the transmitted signal. For each transmon state $\ket{i}$, the spectrum of the resonator typically has a Lorentzian profile at low power, from which we fit $f_{r, \ket{i}}$. The highest state we reach, $\ket{11}$, has a slightly different spectrum, which we attribute to the large charge dispersion affecting our state preparation (see Sec. \ref{sec:coherence}). Each additional transmon excitation red-shifts the resonator frequency such that $\Delta \chi_i \equiv \chi_i - \chi_{i-1} < 0$ for $i < N_{\text{levels}}$, which is consistent with our theoretical prediction. 
The overall amplitude drop in Fig. \ref{fig:3_RFS}(a) is due to our room-temperature electronics, and does not affect our readout method.

From the spectroscopy results above we also find $\sum_i |\Delta \chi_i| > \kappa$. However, this imposes a limit on the amount of transmon states that can be distinguished by a single tone when the total dispersive shift exceeds the resonator linewidth.
For instance, in Fig. \ref{fig:3_RFS}(a), a microwave tone located at frequency $f_a$ is able to distinguish between $\ket{0}$, $\ket{1}$, $\ket{2}$, and $\ket{3}$, but produces similar responses if the transmon is instead in higher excited states like $\ket{8}$, $\ket{9}$, or $\ket{10}$. As a result, a single tone could typically read out at most 4 to 5 states before seeing a significant drop in the readout assignment fidelity.

To overcome this problem we use a multi-tone readout method in which a microwave signal containing multiple simultaneous tones is sent to a single resonator, which has been previously demonstrated for qudit readout \cite{chenTransmonQubitReadout2023, champion2024}. Here we extend this method to include more tones, applying it to a system with much larger dimension and introducing a new analysis of the measured signal. Conceptually, each tone can distinguish only a subset of states, and the information extracted from each of these tones can overlap. For example, suppose a single tone can read out 4 states in the system. We can use three such tones to read out 10 states: the first tone $f_a$ yields one of $\ket{0}$, $\ket{1}$, $\ket{2}$, or $\ket{3+}$, where $\ket{i+}$ ($\ket{i-}$) denotes all states equal to or higher (lower) than $\ket{i}$. The second tone can be placed to distinguish between $\ket{3-}$, $\ket{4}$, $\ket{5}$, and $\ket{6+}$, while a third tone can distinguish between $\ket{6-}$, $\ket{7}$, $\ket{8}$, and $\ket{9}$. Then by combining the results of these three tones we can identify the transmon's state. For other systems, the number of tones and the states they are responsible for reading out will be different depending the system parameters $\Delta \chi_i$ and $\kappa$.
In general, the most important principle of multi-tone readout is to ensure that the transmon can be projected onto one of its eigenstates, and that the state information can be properly detected and carried by at least one of the frequency components in the readout signal.

Multi-tone readout is implemented in our setup using the readout chain shown in Fig. \ref{fig:3_RFS}(c). The multiplexed signal is generated at baseband using a single pair of DACs, then upconverted to RF frequencies and sent to the chip in the dilution refrigerator. After interacting with the resonator, the hanger-transmitted signal \cite{wang2021} will be amplified by a TWPA \cite{macklin2015}, a HEMT low-noise amplifier, then a room-temperature amplifier before being demodulated and sampled by a pair of ADCs, from which we retrieve the raw trace $I[n]$, $Q[n]$. For typical $\Delta \chi_i$ and $\kappa$ in the MHz range, the addition of more readout tones usually does not require more hardware or bandwidth and is compatible with a wide variety of experimental setups. We calibrate our multi-tone readout by placing the frequency of each tone to maximize the readout assignment fidelity for four consecutive transmon states, and the pulse length and amplitude are chosen to balance readout speed and QNDness as shown in Fig. \ref{fig:3_RFS}(d). We note that introducing more tones could increase the average photon number in resonator during readout, depending on the amplitude and phase of each tone, particularly when two tones are close in frequency. Therefore, based on the calibration result, we intentionally reduce the amplitude of the third tone $f_c$ to avoid potential non-Gaussian effects and low QNDness caused by mechanisms such as ionization during readout \cite{shillitoDynamicsTransmonIonization2022}.

\begin{figure}[tp]
\includegraphics{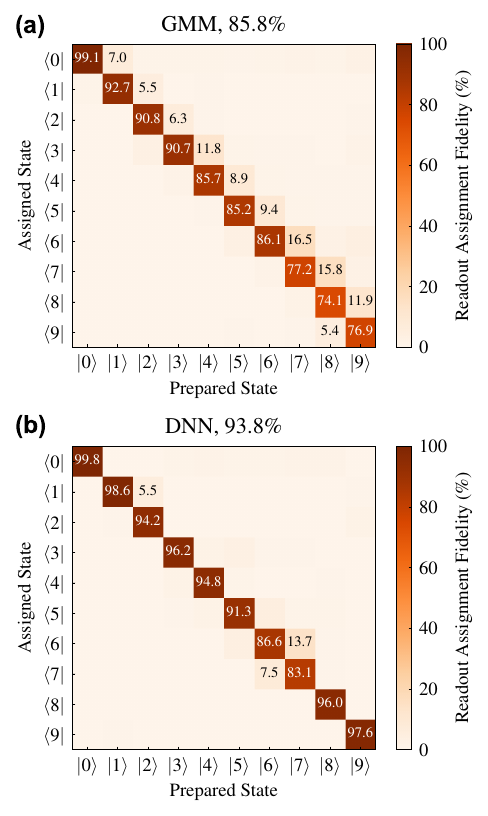}
\caption{
\label{fig:4_fidelity} 
Readout assignment matrices and fidelities for the Gaussian mixture model (a) and deep neural network (b) classification methods. For each transmon state, we acquire 8000 traces for training and validation and an additional 2000 traces for testing the assignment fidelity; the same data is used for both classifiers to ensure that the comparison is fair. Matrix values greater than 5\% are printed on the plots.
}
\end{figure}

After sampling the returned signal, we process the raw trace to produce a transmon state assignment. Here we compare two classification methods: a Gaussian mixture model (GMM) and a deep neural network (DNN). 

In single-tone readout, a single $IQ$ point is obtained via digital demodulation and integration of the raw trace, and is expected to follow a distinct 2D Gaussian distribution in the $I\!-\!Q$ plane for each transmon eigenstate \cite{krantzQuantumEngineerGuide2019}. For multi-tone readout we again obtain a single raw trace, this time containing multiple frequency components. Digital demodulation can likewise be performed at different frequencies, each yielding a different $IQ$ point in each of their respective $I\!-\!Q$ planes. As an example, in Fig. \ref{fig:3_RFS}(b) we show $IQ$ plots for readout of 10 transmon states in three different $I\!-\!Q$ planes. In principle one could do classification in each separate $IQ$ plane, fitting the $IQ$ points with multiple 2D Gaussian distributions before combining the multiple classification results to give one single state assignment. This method, however, has several drawbacks. First, readout errors caused by, e.g., overlap of Gaussians or state transitions during readout can lead to state assignment conflicts. For example, a transmon in state $\ket{4}$ would ideally yield assignments $\ket{3+}$ for the first tone and $\ket{4}$ for the second tone, but readout errors can lead to assignments $\ket{2}$ and $\ket{4}$, which conflict with one another. Second, states not distinguished by a particular tone will not necessarily form a single Gaussian distribution in the $I\!-\!Q$ plane, see Appendix \ref{sec:semi_classical} for more details. Finally, while each individual tone cannot faithfully distinguish between states it is not primarily responsible for reading out, it still carries information from those states that can further assist in classification.

For these reasons, we combine the $IQ$ points of the three tones to form a single six-dimensional vector $(I_a, Q_a, I_b, Q_b, I_c, Q_c)$. For each transmon eigenstate we fit a six-dimensional Gaussian distribution to an ensemble of these points, ultimately yielding a GMM classifier that produces a unique state assignment for each raw trace regardless of the number of tones. We thereby avoid both contradictory classification results and the information loss discussed above. Our GMM is implemented using the \textit{scikit-learn} Python package \cite{pedregosa2011}. 
We calibrate and verify our multi-tone readout method by preparing the transmon $Q_5$ in eigenstates $\ket{0}$ through $\ket{9}$. For each state, we perform 8000 shots of the measurement to produce the GMM training set, followed by an additional 2000 shots for testing the fidelity. In Fig. \ref{fig:4_fidelity}(a) we show the readout assignment matrix, which can also be used to correct population estimates from an ensemble of measurements and mitigate readout errors \cite{kandala2017}. Each element here is a conditional probability $P(\text{assigned as }i| \text{prepared as }j)$, and the readout assignment fidelity $\mathcal{F}$ can be calculated by $\mathcal{F} = \frac{1}{N}\sum^{N-1}_{i=0}P(i|i)$ \cite{gambetta2007}. For the GMM classification method on 10 states we obtain $\mathcal{F}=\readoutfidelityGMM$. 

Although the GMM method is relatively simple to implement and train, the digital demodulation and integration used in acquiring the $IQ$ points will in principle still drop some information present in the full traces $I[n]$ and $Q[n]$. Previous work \cite{lienhardDeepNeuralNetworkDiscriminationMultiplexed2022} has shown that using a DNN as a discriminator that acts directly on the raw traces can effectively reduce assignment errors for multi-qubit readout. Here we extend this method to the case of a single qudit with multiple readout tones. One advantage of a DNN is that the size of the neural network is independent of the number of tones used for readout. For a sampling rate of 1 GS/s and a readout length of \qty{2.2}{\us}, flattening $I[n]$ and $Q[n]$ yields a total of 4400 values per readout shot, corresponding to an input layer with 4400 nodes. Our neural network is fully connected, has two hidden layers with 578 and 76 nodes, respectively, and uses the ReLU as activation function. The output layer has 10 nodes, corresponding to the dimension of our transmon qudit. The DNN is implemented using \textit{pytorch} \cite{paszke2019}, and the Adam optimizer \cite{kingma2017} is used for minimizing the cross-entropy loss function during the training process. In order to ensure a fair comparison, we use same 8000 raw traces for training as were used for the GMM method and 5-fold cross-validation on the same 2000 traces for testing fidelity. The resulting assignment matrix is shown in Fig. \ref{fig:4_fidelity}(b) where we obtain $\mathcal{F}=\readoutfidelityDNN$ for 10 states, a 56\% reduction in assignment error as compared to the GMM method. 

Our results demonstrate that multi-tone readout is an effective means of performing single-shot measurements of high-dimensional qudits. Here we propose several improvements that may further increase the assignment fidelity. First, the dominant errors present in the readout assignment matrix in Fig. \ref{fig:4_fidelity}(a) are the nonzero values of $P(i-1|i)$, which are mostly due to decay during readout. This includes both spontaneous decay and readout-induced decay. The \qty{2.2}{\us} readout time is non-negligible compared to the 10-\qty{20}{\us} $T_1$ lifetimes of the higher levels; see Sec. \ref{sec:coherence}. Using the independently measured $T_1$ results, we estimate that the \qty{2.2}{\us} spontaneous decay can induce up to $3\%$ error for preparing $\ket{1}$ and up to $16\%$ for preparing $\ket{9}$. The readout speed could be further improved by designing larger $\Delta \chi_i$ and $\kappa$ and by adding a Purcell filter to our resonator \cite{reed2010}. Second, we extract the quality factor from the data in Fig. \ref{fig:3_RFS}(a) and find that our resonators are in the under-coupled regime where the internal quality factor is lower than the coupling quality factor, $Q_i < Q_c$ \cite{wang2021}. This causes most readout photons to leak to the environment rather than to the transmission line. This can be improved in future designs by working in the over-coupled regime. In addition, the resonators used in this experiment were coupled to the readout line in hanger-transmission mode. We expect that switching to reflection mode readout could also help to improve the signal-to-noise ratio (SNR) in our system. Finally, although $|\Delta \chi_1|/\kappa \sim 1$ is usually preferred for qubit readout, it is not necessary in the case for transmon qudit readout. We believe carefully designing the dispersive shift while introducing more tones and properly tuning their amplitudes and phases can also improve the assignment fidelity, which we leave for future work.

\section{\label{sec:coherence}Coherence time and charge dispersion}

\begin{figure}[!t]
\includegraphics{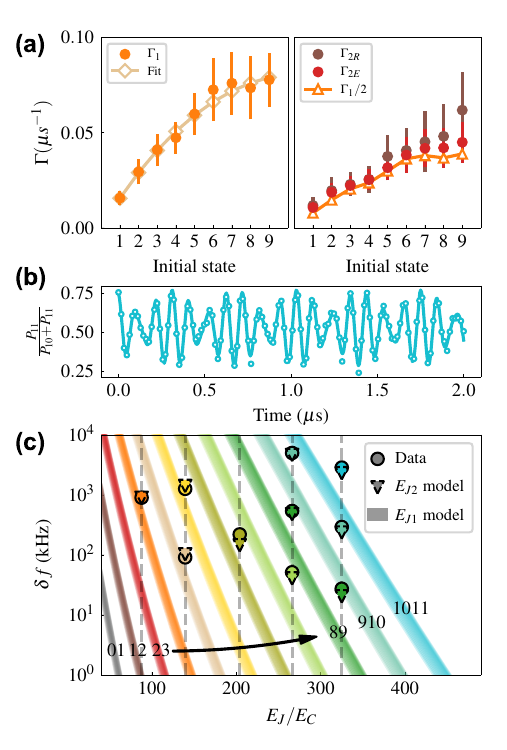}
\caption{
\label{fig:5_coherence} 
Coherence times and charge dispersion. (a) Left: measured relaxation rates $\Gamma_1$ (circles) for each level of $Q_5$. The diamond markers show the fit model explained in Appendix \ref{sec:t1_analysis}. Right: measured dephasing rates $\Gamma_{2R}$ (brown circles) and decoherence rates $\Gamma_{2E}$ (red circles). We plot the relaxation limit $\Gamma_1/2$ for comparison (orange triangles). (b) An example of a Ramsey experiment in the $\{\ket{10},\ket{11}\}$ subspace used for measuring $\delta f$. The normalized population (circles) shows a beating oscillation, which we fit using Eq.(\ref{eq:beating_population}) (solid line). (c) Frequency fluctuations $\delta f$ caused by charge dispersion measured for 5 different transmons and multiple transitions (solid circles). The theoretical predictions given by the standard transmon model Eq. (\ref{eq:charge_dispersion}) are shown in the solid colored bands, where the width of the band is due to the different absolute value of $E_C$ that one could have for the same $E_J/E_C$, which we assume to be \qtyrange{90}{250}{\MHz}. This model yields a lower charge dispersion than the measured value in each case, which we attribute to an overestimate of $E_J$ in our experimental devices: the charge dispersion predicted by the Josephson harmonics model ($M=2$, colored triangles) presented in Sec. \ref{sec: josephson_harmonics} show theoretical corrections for this effect. 
}
\end{figure}

The scaling of coherence times for higher excited states is another important consideration in qudit processors. Fermi's golden rule suggests that the decay rates scale with the matrix elements of the operator associated with the transition process. A simple $LC$ circuit can be modeled as a harmonic oscillator whose annihilation operator goes linearly as $|\bra{i-1} \hat{a} \ket{i}|^2 = i$, which produces the phenomenon known as bosonic enhancement \cite{ExploringtheQuantum}. The transmon differs from the harmonic oscillator in two ways. First, each transition has a different frequency, and we therefore must consider the spectral density of noise separately at each transition. Second, in place of the bosonic operator $\hat{a}$  we must consider the nonlinearity of transmons and use the charge operator $|\bra{i-1} \hat{n} \ket{i}|^2$. Previous work \cite{petererCoherenceDecayHigher2015} has measured transmon coherence times up to $d=5$ and showed good agreement with the Duffing oscillator model. Here we combine the control and readout methods mentioned in previous sections to characterize the coherence of higher levels. 

We quantify our system's coherence properties by performing $T_1$, Ramsey ($T_{2R}$) and Hahn echo ($T_{2E}$) experiments for all adjacent-level transitions on $Q_5$ up to $\ket{9}$. In the $T_1$ experiments we prepare the transmon in an eigenstate $\ket{i}$ and measure the residual population after a variable wait time, extracting $T_1^{i}$ as the time constant of the exponential decay. In the Ramsey and Echo experiments we calculate the normalized population in the $\{\ket{i-1}, \ket{i}\}$ subspace and fit decay constants $T_{2R}^{i-1,i}$ and $T_{2E}^{i-1,i}$. We interleave $T_1$, Ramsey and echo measurements of the excited states for more than 3 days to average over temporal fluctuations. The results are shown in Fig. \ref{fig:5_coherence}(a). Our base transition has $T_1^{1} = \qty{64\pm15}{\us}$, $T_{2R}^{01} = \qty{85\pm31}{\us}$, and $T_{2E}^{01} = \qty{93\pm27}{\us}$, while the highest transition has $T_1^{9} = \qty{13\pm2}{\us}$, $T_{2R}^{89} = \qty{16\pm5}{\us}$, and $T_{2E}^{89} = \qty{22\pm5}{\us}$. We find that our measured echo times $T_{2E}$ approach $2T_{1}$, especially for the higher transitions, suggesting that there are no additional strong dephasing channels and that the coherence is mostly limited by $T_1$ decay. 
The pure dephasing time $T_{\phi}$ can be estimated through $1/T_{\phi} \approx 1/T_{2E} - 1/(2T_1)$ \cite{krantzQuantumEngineerGuide2019}, which we found to be around $\qty{300}{\us}$ in $\{\ket{0}, \ket{1}\}$ subspace and $\qty{200}{\us}$ in $\{\ket{8}, \ket{9}\}$ subspace. 
In general there are multiple noise sources that can induce transmon relaxation, including Purcell decay, quasiparticle tunneling, dielectric loss and others. Each noise source scales differently with increasing level and has a different frequency dependence. We give a detailed analysis in Appendix \ref{sec:t1_analysis}. Overall, the higher levels of the transmon have shorter $T_1$ times due to bosonic enhancement, where the strongest noise channel in our system is likely dielectric loss.

In addition to the coherence time, charge dispersion is an important factor to consider when assessing the viability of the transmon as a high-dimensional qudit. As illustrated in Sec. \ref{sec:concept}, we expect the high-$E_J/E_C$ transmon to suppress the charge dispersion $\epsilon_m$ of the higher energy levels. Here we measure the frequency fluctuation $\delta f = (|\epsilon_m| + |\epsilon_{m+1}|) / h$ on different transmons spanning a wide range of $E_J/E_C$ values. The effective offset charge $n_g$ can be changed by two mechanisms \cite{serniak2018}: quasiparticle tunneling across the Josephson junction and reconfiguration of mobile charges in the environment. Quasiparticle tunneling has a time scale of around a millisecond \cite{serniak2018, tennant2022, pan2022}, and has the effect of changing $n_g$ by $\pm 1/2$, which switches the charge parity of transmon and causes the transition frequency to jump between two values $f_e$ and $f_o$. Each individual shot of the Ramsey experiment takes only $\sim \qty{10}{\us}$, far less than the quasiparticle tunneling time scale, and therefore samples only a single realization of the quasiparticle-shifted transition frequency. The measured oscillations in the population are a result of averaging over many realizations of this experiment over the course of roughly 10 seconds, which leads to a beating between the different oscillation frequencies $f_e$ and $f_o$,  as shown in Fig. \ref{fig:5_coherence}(b).
Environmental charge reconfiguration, meanwhile, occurs on a timescale of minutes and can change $n_g$ by an arbitrary amount, thereby changing the beat frequency. 
Hence, such a Ramsey experiment repeated for several hours can be used to extract $\delta f = \text{max} \; |f_o - f_e|$ from fluctuations of the frequency difference $f_o - f_e$ due to a varying $n_g$ \cite{serniak2018}. 

In principle, the method described above could be applied to any transition in transmon. In practice, however, relaxation of the transmon limits the durations of the Ramsey scans, inevitably resulting in a finite frequency resolution. Even though the lower transitions have relatively longer lifetimes, their extremely small frequency fluctuations ($\delta f < \qty{20}{\kHz}$) are still below the available resolution, thus cannot be measured by this method.
For each of the other transitions, we normalize the population within the Ramsey subspace and fit the resulting oscillations with
\begin{equation}
\begin{split}
P_{\text{norm}}(t) =  C + e^{-t/T_{2R}} [&A_0 \cos(2\pi f_e t + \phi_0) \\
                                       + &A_1 \cos(2\pi f_o t + \phi_1)],
\label{eq:beating_population}
\end{split}
\end{equation}
where $C$, $T_{2R}$, $A_0$, $A_1$, $f_e$, $f_o$, $\phi_0$, and $\phi_1$ are the fitting parameters. Notably, we treat $A_0$ and $A_1$ as independent parameters because the transmon can stay in the odd- or even-parity states for different amounts of time. We typically fix $\phi_0=\phi_1=0$ for the lower transitions, but for large frequency fluctuations ($\delta f > \qty{2}{\MHz}$), the resulting detuning of the $\pi/2$ pulse will cause the preparation of the equal superposition state to be imperfect. We account for this by leaving the two phases as free parameters.

The measured $\delta f$ is shown in Fig. \ref{fig:5_coherence}(c), where we also include the numerical result \cite{groszkowski2021} predicted by the standard transmon Hamiltonian. The frequency fluctuations $\delta f$ of the higher transitions of the high-$E_J/E_C$ transmons are comparable with the $\delta f$ values of the lower transitions for common low-$E_J/E_C$ transmons. For example, the $\ket{3} \leftrightarrow \ket{4}$ transition of $Q_0$ ($E_J/E_C = 88$) has $\delta f = \qty{901}{\kHz}$ whereas the $\ket{9} \leftrightarrow \ket{10}$ transition of $Q_5$ ($E_J/E_C = 325$) only has $\delta f = \qty{292}{\kHz}$. This validates the idea that high values of $E_J/E_C$ effectively suppress charge dispersion in the highly excited states, extending the concept underlying the standard transmon beyond its lowest few eigenstates. The theoretical prediction of $\delta f$ depends not only on $E_J/E_C$ but also on the absolute value of $E_C$ (or equivalently, absolute frequency $f_{01}$). We therefore plot it as a solid band spanning common values of $E_C$ in Fig. \ref{fig:5_coherence}(c). We find, however, that the standard transmon model predicts $\delta f$ smaller than our measured results. As discussed in Sec. \ref{sec:model}, we find that this discrepancy can be reduced by considering the Josephson harmonics model \cite{willschObservationJosephsonHarmonics2024}.

\section{\label{sec:model}Verification of the transmon model}

\subsection{\label{sec: josephson_harmonics}Josephson harmonics}

The standard transmon Hamiltonian, Eq. (\ref{eq:standard_transmon_hamiltonian}), has proven to be an effective model for transmons operated as qubits and qutrits. Recent work \cite{willschObservationJosephsonHarmonics2024} on Josephson harmonics, however, showed that a more detailed current-phase relation can lead to a correction to the transmon Hamiltonian that yields a better model for the systems studied. Such an improvement can be important for high-$E_J/E_C$ transmons for several reasons. First, the standard transmon model, with its two free parameters $E_J$ and $E_C$, can satisfactorily predict $f_{01}$ and $f_{12}$, but usually begins to diverge from measurements for higher transitions. This is important for transmons with high $E_J / E_C$ because it is only in this case that these levels are stable and usable. Second, when we design the system for a target dimension $d$, the standard transmon Hamiltonian may underestimate the charge dispersion, as shown in Fig. \ref{fig:5_coherence}(c), causing unexpectedly short coherence times for the higher transitions. Furthermore, with more transition frequencies in the system, the calibration routine of a high-$E_J/E_C$ transmon requires additional effort. A better prediction of the system parameters can help to reduce the time to calibrate a large-scale system. Previous work \cite{willschObservationJosephsonHarmonics2024} has explored 7 levels in transmons. Here we apply the Josephson harmonics model to our system up to 12 levels. Our high-fidelity control and readout methods, along with our relatively long coherence times on the higher levels, make our system an ideal platform to verify and utilize such a correction.

\begin{figure}[tp]
\includegraphics{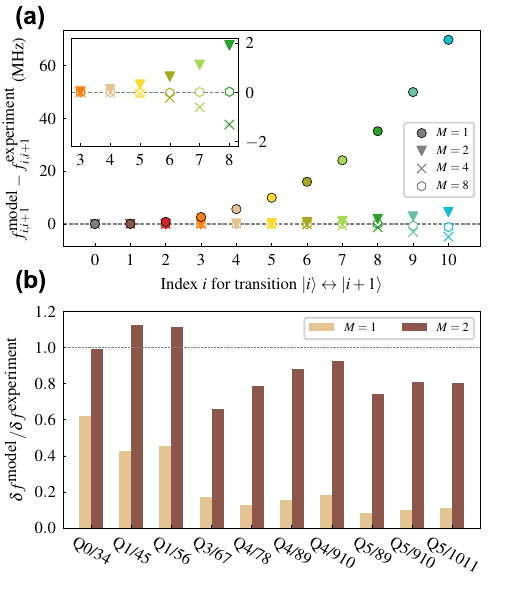}
\caption{
\label{fig:6_harmonics} 
Transmon model with Josephson harmonics. (a) Difference between the numerical prediction and experimental result for each transition frequency $f_{i,i+1}$ on $Q_5$. The frequencies predicted by the standard transmon model ($M=1$) have larger discrepancies than those predicted by the Josephson harmonics models ($M=2,4,8$). (b) Comparison of the predicted frequency fluctuation $\delta f^{\text{model}}$. The transmon harmonics model ($M=2$) shows better agreements with experimental results (dashed line) than the standard transmon model ($M=1$).
}
\end{figure}

For a transmon capacitively coupled with a linear resonator, the Josephson harmonics model truncated at the $M$-th harmonic ($E_{JM}$ model) gives Hamiltonian
\begin{equation}
\begin{split}
\hat{H}_{\text{Jh}} & = 4 E_C (\hat{n}-n_g)^2 - E_{J1} \cos\hat{\phi}  \\
  & - \sum_{m\geq2}^M E_{Jm}\cos(m\hat{\phi}) + \hat{H}_r + \hat{V}_{t-r}.
\label{eq:harmonic_transmon_Hamiltonian}
\end{split}
\end{equation}
The $E_{J1}$ model drops all $E_{Jm}$ terms and recovers the standard transmon-resonator Hamiltonian. The Josephson harmonic energies $E_{Jm}$ can be predicted by a physical model which suggests that they should alternate in sign and decrease in magnitude for increasing $m$ \cite{willschObservationJosephsonHarmonics2024}. In this model, a stronger inhomogeneity in the barrier thickness of the Josephson junction leads to a larger magnitude of $E_{Jm}$.

We begin our evaluation of the different $E_{JM}$ models by considering the transition frequencies of our high-$E_J/E_C$ transmon. These frequencies, initially determined by spectroscopy, are further calibrated to the tens of kilohertz level using Ramsey experiments. We use the frequencies of the lower transitions to fit the parameters in Eq. (\ref{eq:harmonic_transmon_Hamiltonian}), then compare the predictions for the higher transition frequencies to the experimental measurements. To reduce underfitting and overfitting we use the lowest $M+1$ transition frequencies to fit the $E_{JM}$ model. Transitions higher than $\ket{8} \leftrightarrow \ket{9}$ are not used in fitting due to their large charge dispersions. For example, the measured $f_{01}$, $f_{12}$, $f_{23}$, $f_{r,\ket{0}}$, an $f_{r,\ket{1}}$ values are used to fit $E_{J1}$, $E_{J2}$, $E_C$, $f_r$, and $g$ in the $E_{J2}$ model. We enforce $E_{Jm}$ to alternate in sign during fitting, and the full fitting results can be found in Appendix \ref{sec:device_params}.

We show the results for each transition frequency of $Q_5$ in Fig. \ref{fig:6_harmonics}(a), comparing the differences between the numerical predictions given by the model, $f_{i, i+1}^{\text{model}}$, and the experimentally measured frequencies $f_{i, i+1}^{\text{experiment}}$. We emphasize that this comparison can only be made for the transitions whose frequencies are not used in fitting. The standard transmon model ($M=1$) uses $f_{01}$ and $f_{12}$ for fitting, and its prediction begins to deviate at $f_{23}$. This is followed by a superlinear increase to tens of megahertz for the higher transitions. The second simplest model ($M=2$), meanwhile, reduces the discrepancy by an order of magnitude despite having only one more free parameter in the Hamiltonian. Moreover, we find that introducing more harmonics (i.e., increasing $M$) in Eq. (\ref{eq:harmonic_transmon_Hamiltonian}) further improves the precision of the frequency predictions, which agrees with the physical intuition that the true junction potential has infinitely many harmonics, $M \rightarrow +\infty$.

In Fig. \ref{fig:6_harmonics}(b) we compare the the frequency fluctuations $\delta f^{\text{model}}$ predicted by the standard transmon ($E_{J1}$) model and the $E_{J2}$ model to the experimental result $\delta f^{\text{experiment}}$. We find that the result given by the $E_{J2}$ model better predicts $\delta f$. The standard transmon model underestimates $\delta f$ in our system, likely because it overestimates $E_{J1}/E_C$; this deviation is accounted for by the Josephson harmonics model. As an example, the standard transmon model estimates $E_{J1}/E_C=325$ on $Q_5$, whereas all Josephson harmonics models estimate $270<E_{J1}/E_C<290$. This overestimation becomes more apparent as $E_J/E_C$ is increased, as can be seen in Fig. \ref{fig:5_coherence}(c). Combining this with the observation that transmons with larger values of $E_{J1}$ typically have larger harmonic ratios $|E_{J2}/E_{J1}|$ on our devices (see Appendix \ref{sec:device_params}), we hypothesize that transmons with larger values of $E_{J1}$ are affected more by the Josephson harmonic effect. We also find that including more harmonics does not further improve the estimation of $\delta f$, due to their relatively small corrections to the $E_{J1}$. We note that while the correction on $E_{J1}/E_C$ will in principle change the absolute value of the charge matrix elements $|\bra{i} n \ket{i+1} | ^ 2 $, it will not significantly affect their relative magnitudes, leaving our $T_1$ fitting in Sec. \ref{sec:coherence} largely unaffected.

\subsection{\label{sec: zz_coupling}$ZZ$ coupling}

\begin{figure}[bp]
\includegraphics{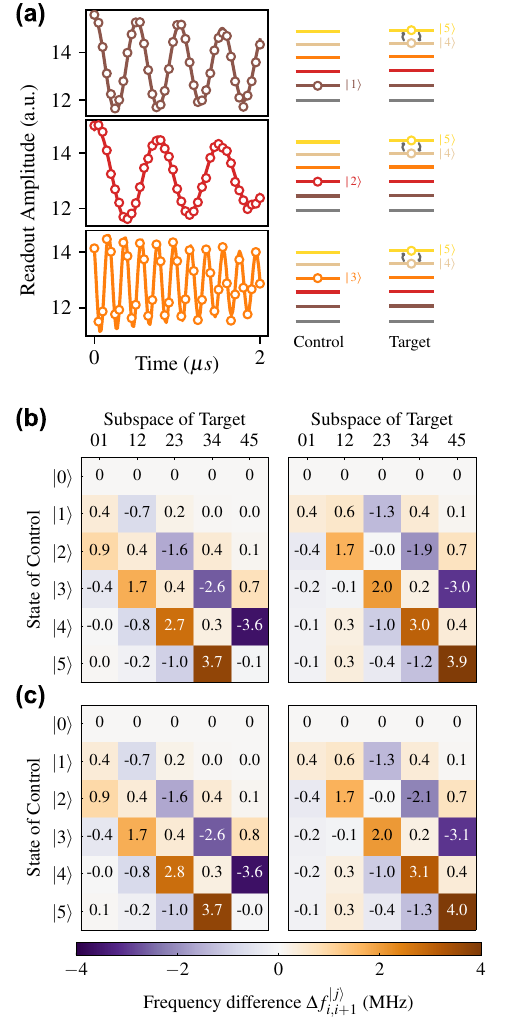}
\caption{
\label{fig:7_coupling} 
Effective $ZZ$ coupling strength between two high-$E_J/E_C$ transmons. (a) Example of a Ramsey experiment in the $\{\ket{4},\ket{5}\}$ subspace of $Q_3$ (target) when $Q_2$ (control) is in $\ket{1}$, $\ket{2}$ and $\ket{3}$. The $ZZ$ coupling has the effect of inducing control state-dependent shifts on the target transmon's transition frequencies, which are accurately resolved by the Ramsey experiments. (b) Measured frequency shifts of the target transmon's transitions for each control state. In the left plot we assign $Q_3$ the role of the control while $Q_2$ is the target, and in the right plot we swap these roles. Each value reported here is the difference in frequency of the indicated target transition between when the control is in the indicated state and when it is in $\ket{0}$. 
(c) Numerical calculations of the frequency shifts measured in (b). The native coupling strength $J$ is fitted from $\Delta f^{\ket{1}}_{01}$.
}
\end{figure}

Having characterized control, single-shot readout, and coherence in our high-$E_J/E_C$ transmon qudits, we now turn our attention to two-qudit interactions. Our platform provides a unique opportunity to examine the behavior of transmon-transmon coupling in highly excited states. Here we consider direct capacitive coupling between $Q_1$ and $Q_2$ on the chip B, see Appendix \ref{sec:device_params}. The Hamiltonian of this system is 
\begin{equation}
\hat{H}^{(12)}_{t-t} = \hat{H}^{(1)}_{\text{t-r}} + \hat{H}^{(2)}_{\text{t-r}} + hJ\hat{n}^{(1)}\hat{n}^{(2)}
\label{eq:standard_transmon_transmon_hamiltonian}   
\end{equation}
where $J$ is the transmon-transmon coupling strength. 
In the dispersive limit where this coupling strength is small relative to the detuning between the transmons, the last term in Eq. (\ref{eq:standard_transmon_transmon_hamiltonian}) amounts to an effective $ZZ$-like coupling in which each state of one transmon (the control) induces a small frequency shift on the transitions of the other transmon (the target). These frequency shifts can be detected with Ramsey experiments, which we demonstrate in Fig. \ref{fig:7_coupling}(a).

We design $Q_1$ and $Q_2$ to have similar values for $E_C$ but slightly different values for $E_J$ such that all of their transition frequencies are properly detuned, $\Delta_{ij} = \text{min}_{ij} |f^{(1)}_{i,i+1} - f^{(2)}_{j,j+1}| \gg J$, thereby avoiding frequency collisions, see Appendix \ref{sec:device_params}. We show the measured frequency shifts in Fig.\ref{fig:7_coupling}(b) where we consider the lowest 6 levels for both transmons and also switch the role of control and target. Each value in the matrix is the difference in frequency of a target transition between when the control is in $\ket{j}$ and when it is in $\ket{0}$, which we denote as $\Delta f^{\ket{j}}_{i,i+1}$ where $\Delta f^{\ket{0}}_{i,i+1}=0$. Although the native coupling strength $J$ is fixed in our system, the effective $ZZ$-like coupling is boosted by bosonic enhancement-like increase in the charge operators' matrix elements. For instance, if we encode $Q_1$ (control) and $Q_2$ (target) as qubits ($d=2$) in their $\{\ket{0},\ket{1}\}$ subspaces, the frequency shift is $\Delta f^{\ket{1}}_{01} - \Delta f^{\ket{0}}_{01} \simeq \qty{0.4}{\MHz}$. If the qubits are encoded in $\{\ket{4},\ket{5}\}$ subspace, however, the effective $ZZ$ strength is $\Delta f^{\ket{5}}_{45} - \Delta f^{\ket{4}}_{45} \simeq \qty{3.5}{\MHz}$. In general, the frequency shifts induced by the coupling are not only scaled by the absolute coupling strength $J$, but also by the charge operator matrix elements, which grow as the control and target transmons move to higher levels.

In Fig.\ref{fig:7_coupling}(c) we show a theoretical calculation for these frequency shifts obtained by numerically diagonalizing the two-transmon Hamiltonian, Eq.(\ref{eq:standard_transmon_transmon_hamiltonian}). Here we fit $J$ from $\Delta f^{\ket{1}}_{01}$ alone. We find that our theoretical prediction agrees very well with the experimental results. Due to the similar parameters between the two transmons $Q_1$ and $Q_2$, including the Josephson harmonics model will not significantly change their relative detunings bewteen same transitions and thus will not strongly impact the results here. The Josephson harmonics model can be more helpful for a pair of transmons with distinct $E_J/E_C$ and anharmonicities. The ability to accurately model the interaction in theory along with the strong coupling of higher energy levels suggests a path towards fast two-qudit gates and two-qubit gates with enhanced interaction rate, which we will leave for future work.

\section{\label{sec:conclusion}Conclusions and Outlook}

In this work we have demonstrated the feasibility of operating high-$E_J/E_C$ transmons as a qudits. We observed the existence of 12 transmon levels and demonstrated high-fidelity control in a large Hilbert space up to $d=10$. Our multi-tone method enables efficient single-shot readout of all these levels and yields a \readoutfidelityDNN\ 10-state readout assignment fidelity, with further achievable improvement left for future work. Our $T_1$ and $T_2$ measurements of the higher levels show stable coherence, in particular demonstrating that charge dispersion is effectively suppressed by modifying the standard transmon design to realize a larger $E_J/E_C$ ratio. We find that the transition frequencies of the higher levels deviate from those predicted by the standard transmon model, and furthermore show that these deviations can be accounted for by considering the Josephson harmonics. Finally, we show that the effective $ZZ$-like coupling strength in a dispersively coupled transmon pair is boosted by bosonic enhancement in the higher levels, showing close agreement with theoretical predictions.

This work systematically investigates the higher energy levels of transmons and establishes the high-$E_J/E_C$ transmon as a platform for high-dimensional quantum information processing. Our chip design and experimental setup follow the conventional circuit QED architecture, and do not introduce strong additional hardware requirement. The system may be scaled up with proper solutions for frequency crowding \cite{theis2016, krinner2020}. In the future, with novel fabrication materials and methods \cite{place2021, wang2022}, the coherence time of our system could be further improved for all levels. The large $ZZ$ coupling strength also provides a possible opportunity for creating a fast qudit entangling gate \cite{goss2022}. Furthermore, with access to higher transmon states, our system could also be used for testing and verifying complicated transmon dynamics such as readout-induced state transitions \cite{shillitoDynamicsTransmonIonization2022, khezri2023}.

\section*{Acknowledgments}

This work is spported by Air Force Office of Scientific Research under award number FA9550-23-1-0121. Devices used in this work were fabricated and provided by the Superconducting Qubits at Lincoln Laboratory (SQUILL) Foundry at MIT Lincoln Laboratory, with funding from the Laboratory for Physical Sciences (LPS) Qubit Collaboratory.The traveling-wave parametric amplifier (TWPA) used in this experiment was provided by IARPA and Lincoln Labs.

\appendix

\section{\label{sec:device_params}Device parameters}

\begin{table*}[tp]

\caption{\label{tab:device_params}Device parameters.}

\begin{ruledtabular}

\begin{tabular}{ccccccc}

Device & $Q_0$ & $Q_1$ & $Q_2$ & $Q_3$ & $Q_4$ & $Q_5$ \\ 

\hline

Chip  & Chip A & Chip B & Chip B & Chip C & Chip C & Chip C \\

Josephson energy $E_J/h$ (GHz) \footnote{Parameters here is estimated by standard transmon model.} & 16.685 & 21.194 & 21.960 & 25.636 & 28.702 & 32.191 \\

Charging energy $E_C/h$ (GHz) \footnotemark[1] & 0.190 & 0.152 & 0.152 & 0.126 & 0.108 & 0.099 \\

$E_J/E_C$ \footnotemark[1] & 88 & 139 & 144 & 204 & 266 & 325 \\ 

\hline

First anharmonicity $\alpha_1 = f_{12} - f_{01}$ (MHz) & -209 & -164 & -164 & -133 & -113 & -104 \\ 












0-1 transition frequency $f_{01}$ (GHz) & 4.8365 & 4.9177 & 5.0143 & 4.9449 & 4.8640 & 4.9472 \\

1-2 transition frequency $f_{12}$ (GHz) & 4.6272 & 4.7541 & 4.8505 & 4.8117 & 4.7506 & 4.8437 \\

2-3 transition frequency $f_{23}$ (GHz) & 4.3992 & 4.5794 & 4.6771 & 4.6713 & 4.6317 & 4.7356 \\

3-4 transition frequency $f_{34}$ (GHz) & 4.1463 & 4.3912 & 4.4898 & 4.5223 & 4.5066 & 4.6225 \\

4-5 transition frequency $f_{45}$ (GHz) & - & 4.1859 & 4.2862 & 4.3630 & 4.3743 & 4.5036 \\

5-6 transition frequency $f_{56}$ (GHz) & - & - & - & 4.1911 & 4.2335 & 4.3779 \\

6-7 transition frequency $f_{67}$ (GHz) & - & - & - & 3.9978 & 4.0822 & 4.2444 \\

7-8 transition frequency $f_{78}$ (GHz) & - & - & - & - & 3.9181 & 4.1015 \\

8-9 transition frequency $f_{89}$ (GHz) & - & - & - & - & 3.7374 & 3.9468 \\

9-10 transition frequency $f_{9,10}$ (GHz) & - & - & - & - & 3.5353 & 3.7772 \\

10-11 transition frequency $f_{10,11}$ (GHz) & - & - & - & - & - & 3.5886 \\

\hline

Resonator frequency when transmon is at $\ket{0}$ $f_{r, \ket{0}}$ (GHz) & 6.682017 & 6.773781 & 6.804014 & 6.377927 & 6.412255 & 6.468937 \\

Dispersive shift $f_{r, \ket{1}} - f_{r, \ket{0}}$ (kHz) & -259 & -228 & -252 & -355 & -201 & -265 \\

Transmon-resonator coupling strength $g$ (MHz) & 33.1 & 31.0 & 31.3 & 30.9 & 25.0 & 28.1 \\

Transmon-transmon coupling strength $J$ (MHz) & - &\multicolumn{2}{c}{1.59} & - & - & - \\

\hline

$E_{J2}/E_{J1}$ \footnote{The result here is fitted using Josephson harmonic model with $M=2$.} & -0.31\% & -0.46\% & -0.05\% & -0.53\% & -0.62\% & -0.66\% \\ 

$f_{34, \text{num}} - f_{34, \text{exp}}$ \footnotemark[2] (kHz) & -441.8 & -219.1 & 2153.4 & 25.5 & 6.3 & 41.9 \\

\end{tabular}

\end{ruledtabular}

\end{table*}

\begin{table*}[tp]

\caption{\label{tab:josephson_harmonics_params}$Q_5$ Parameters of Josephson harmonics models. All parameters are given in gigahertz.}

\begin{ruledtabular}

\begin{tabular}{cccccccccccc}

Model & $E_C/h$ & $E_{J1}/h$ & $E_{J2}/h$ & $E_{J3}/h$ & $E_{J4}/h$ & $E_{J5}/h$ & $E_{J6}/h$ & $E_{J7}/h$ & $E_{J8}/h$ & $f_r/h$ & $g/h$ \\ 

\hline

standard & 0.099 & 32.1906 & - & - & - & - & - & - & - & 6.468 & 0.028 \\

$E_{J2}/h$ & 0.107 & 30.7166 & -0.2025 & - & - & - & - & - & - & 6.468 & 0.029 \\

$E_{J3}/h$ & 0.111 & 30.1116 & -0.3698 & 0.0161 & - & - & - & - & - & 6.468 & 0.030 \\

$E_{J4}/h$ & 0.109 & 30.3566 & -0.3303 & 0.0176 & -0.0014 & - & - & - & - & 6.468 & 0.029 \\

$E_{J5}/h$ & 0.106 & 30.7248 & -0.1937 & 0.0007 & -0.0016 & 0.0005 & - & - & - & 6.468 & 0.029 \\

$E_{J6}/h$ & 0.108 & 30.5029 & -0.2856 & 0.0138 & -0.0022 & 0.0003 & -0.0000 & - & - & 6.468 & 0.029 \\

$E_{J7}/h$ & 0.107 & 30.6283 & -0.2128 & 0.0000 & -0.0017 & 0.0020 & -0.0011 & 0.0003 & - & 6.468 & 0.029 \\

$E_{J8}/h$ & 0.109 & 30.4054 & -0.3409 & 0.0251 & -0.0038 & 0.0004 & -0.0004 & 0.0005 & -0.0002 & 6.468 & 0.029 \\ 

\end{tabular}

\end{ruledtabular}

\end{table*}

The device parameters used in this work are listed in Table.\ref{tab:device_params}. The frequencies reported here are experimental results and could be affected by temporal fluctuation and large charge dispersions. Some transition frequencies are not measured and reported for low-$E_J/E_C$ transmons due to their limited accessibility. Chip A and C are designed to have negligible transmon-transmon coupling. The chips are mounted at the \qty{10}{\milli\K} stage in a dilution refrigerator. Our control and readout pulse is generated and detected at room temperature using Qblox instrument with sample rate 1 GS/s.

The parameters of Josephson harmonics model is shown in Table.\ref{tab:josephson_harmonics_params}. Previous work \cite{willschObservationJosephsonHarmonics2024} shows that these Josephson harmonic energies $E_{Jm}$ of a tunnel junction should decrease in magnitude for higher order $m$. We observe the same general trend of our fitting results.

\section{\label{sec:state_tomography}Gate set of qudit state tomography}

We reconstruct the density matrix and calculate the state fidelity through qudit state tomography. A proper gate set for the qudit state tomography should be able to transform the off diagonal elements onto the diagonal such that it can be measured by our dispersive readout, which is ideally a projective measurement. Besides the identity gate $I$ which measures the diagonal elements, the other gates are designed by following rules: for the element $\rho_{i, i+1}$, we will use $X^{i,i+1}_{\pi/2}$ and $Y^{i,i+1}_{\pi/2}$ to map its real and imaginary parts to the diagonal; for element $\rho_{i, i+k}$, it will be first populated down to $\rho_{i, i+1}$ by $X^{i+k-1,i+k}_{\pi}$, $X^{i+k-2,i+k-1}_{\pi}$,..., $X^{i+1,i+2}_{\pi}$, and then be mapped to diagonal by $X^{i,i+1}_{\pi / 2}$ and $Y^{i,i+1}_{\pi / 2}$. For example, the gate set we used for $d=3$ is
\begin{equation}
G = \{ I, X^{01}_{\pi/2}, Y^{01}_{\pi/2}, X^{12}_{\pi/2}, Y^{12}_{\pi/2}, X^{01}_{\pi/2}X^{12}_{\pi}, Y^{01}_{\pi/2}X^{12}_{\pi} \}.
\label{eq:qutrit_gate_set}   
\end{equation}
For $d=4$, the gate set is
\begin{equation}
\begin{split}
G = \{ & I, X^{01}_{\pi/2}, Y^{01}_{\pi/2}, X^{12}_{\pi/2}, Y^{12}_{\pi/2}, X^{23}_{\pi/2}, Y^{23}_{\pi/2}, \\
& X^{01}_{\pi/2}X^{12}_{\pi}, Y^{01}_{\pi/2}X^{12}_{\pi}, 
X^{12}_{\pi/2}X^{23}_{\pi}, Y^{12}_{\pi/2}X^{23}_{\pi}, \\
& X^{01}_{\pi/2}X^{12}_{\pi}X^{23}_{\pi}, Y^{01}_{\pi/2}X^{12}_{\pi}X^{23}_{\pi} \}.
\label{eq:ququart_gate_set}  
\end{split}
\end{equation}

We note that our gate set is just one of the possible choices for qudit state tomography. It has the benefit of conceptual simplicity and ease of computation. Other tomography gate sets can be used for reducing the number of measurements or reducing the impact of certain error channels \cite{thew2002, bianchettiControlTomographyThree2010, cramer2010, yurtalan2020, stricker2022, nguyenEmpoweringHighdimensionalQuantum2023}.

\section{\label{sec:dispersive}Dispersive readout}

\begin{figure*}[!t]
\includegraphics{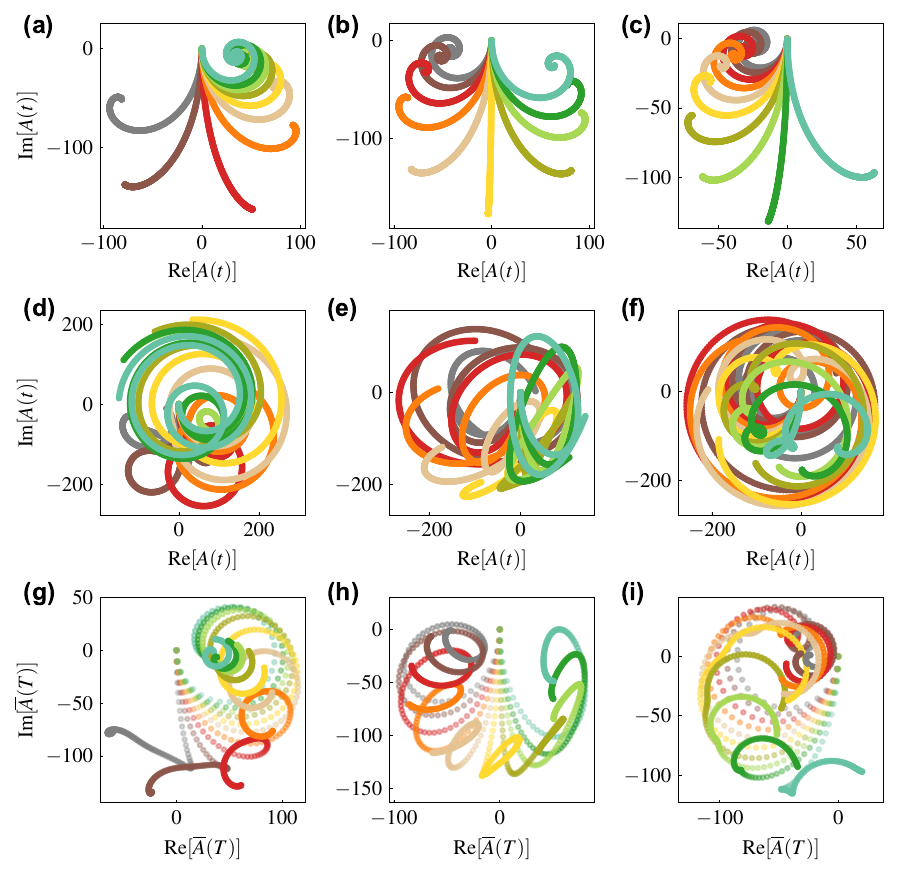}
\caption{
\label{fig:8_semiclassical}
Semi-classical dynamics of a driven resonator in a $I\!-\!Q$ plane. The experimental values of $Q_5$ are used in the calculation. The transmon states are differentiated by colors using same convention as Fig.\ref{fig:3_RFS}(a). (a-c) Trajectories of $A(t)$ in Eq.(\ref{eq:semi_classical_solution}) for the single-tone readout at frequency (a) $\omega_d=\omega_m=\omega_1$, (b) $\omega_d=\omega_m=\omega_2$, (c) $\omega_d=\omega_m=\omega_3$. (d-f) Trajectories of $A(t)$ for the three-tone frequency-multiplexing readout at $\omega_1, \omega_2, \omega_3$ with rotating frequency (d) $\omega_m=\omega_1$, (e) $\omega_m=\omega_2$, (f) $\omega_m=\omega_3$. (g-i) Integrated amplitude $\overline{A}(T)$ for the three-tone frequency-multiplexing readout at $\omega_1, \omega_2, \omega_3$ with rotating frequency (g) $\omega_m=\omega_1$, (h) $\omega_m=\omega_2$, (i) $\omega_m=\omega_3$.
}
\end{figure*}

\subsection{\label{sec:SWT}Dispersive Hamiltonian}

Following the method in \cite{blaisCircuitQuantumElectrodynamics2021, zhu2013}, Eq. (\ref{eq:standard_dispersive_hamiltonian}) can be obtained by performing Schrieffer-Wolff transformation on the Hamiltonian in Eq. (\ref{eq:standard_transmon_resonator_hamiltonian}) where we treat $\hat{V}_{t-r}$ as a perturbation. To the second order, the $\tilde{f}_i$ and $\chi_i$ in Eq. (\ref{eq:standard_dispersive_hamiltonian}) can be expressed by
\begin{equation}
\tilde{f}_i = f_i + \sum_{i'}\chi_{ii'},
\label{eq:lamb_shift}
\end{equation}
\begin{equation}
\chi_i = \sum_{i'}(\chi_{ii'}-\chi_{i'i}),
\label{eq:acstark_shift}
\end{equation}
\begin{equation}
\chi_{ii'} = \frac{g^2 |\bra{i} \hat{n} \ket{i'}|^2}{f_i - f_{i'} - f_r}.
\label{eq:chi_definition}
\end{equation}

\subsection{\label{sec:semi_classical}Semi-classical dynamics of multi-tone readout}

We provide a simplified semi-classical model of a driven resonator following the method given in \cite{kehrerImprovingTransmonQudit2024} to further illustrate our multi-tone readout. We emphasize this only works in the dispersive regime where the existence of transmon only provides a frequency shift to the resonator, and we ignore any readout-induced state transition of the transmon. For a more detailed description of the readout dynamics, we refer to \cite{dumas2024}.

For convenience, we use angular frequencies $\omega = 2 \pi f$ in this section. We denote the complex resonator frequency as $\overline{\omega}_{r, \ket{j}} =\omega_{r, \ket{j}} - \text{i} \kappa / 2$ when transmon is at state $\ket{j}$, and denote the readout frequency of each tone as $\omega_d$. In a frame rotating at the frequency $\omega_m$, the equation of motion of the mean-field amplitude $A(t)$ is
\begin{equation}
\dot{A}(t) = - \text{i} (\overline{\omega}_{r, \ket{j}} - \omega_m) A(t) - \sum_{d=1}^{D} \text{i}\frac{\Omega_d}{2} e^{-\text{i} (\omega_d - \omega_m) t - \text{i} \phi_d},
\label{eq:semi_classical_EOM}
\end{equation}
where $\Omega_d$ and $\phi_d$ are the driving amplitude and phase of each readout tone, and $D$ is the number of tones. With the initial condition that the resonator is in the vacuum state $A(0)=0$, the solution of Eq. (\ref{eq:semi_classical_EOM}) is 
\begin{equation}
A(t) = \sum_{d=1}^{D} \frac{\Omega_d}{2} e^{-\text{i} (\omega_d - \omega_m) t - \text{i} \phi_d} \frac{e^{- \text{i} (\overline{\omega}_{r, \ket{j}} - \omega_d) t} - 1}{\overline{\omega}_{r, \ket{j}} - \omega_d}.
\label{eq:semi_classical_solution}
\end{equation}

The amplitude $A(t)$ shows classical dynamics of driven resonator. For the case of low internal loss, $\kappa_i \ll \kappa_c$, the signal leaking from the resonator will mostly go into the transmission line, be amplified at multiple stages, and finally be detected by the room temperature electronics. The averaged integrated $IQ$ value of $P$ shots measurements can be related to the amplitude $A(t)$ through
\begin{equation}
\begin{split}
\overline{I} &= \sum_{p}^P I_p = \sum_{p=1}^P \sum_{n} e^{-\text{i}\omega_m t_n}I_p[n] \overset{\mathrm{P\rightarrow\infty}}{\longrightarrow} \text{Re}[\overline{A}(T)], \\
\overline{Q} &= \sum_{p}^P Q_p = \sum_{p=1}^P \sum_{n} e^{-\text{i}\omega_m t_n}Q_p[n] \overset{\mathrm{P\rightarrow\infty}}{\longrightarrow} \text{Im}[\overline{A}(T)],
\label{eq:semi_classical_IQ}
\end{split}
\end{equation}
where
\begin{equation}
\overline{A}(T) = \int_0^T A(t)\text{d}t,
\label{eq:semi_classical_integrated_amplitude}
\end{equation}
and $T$ is the integration length. In Fig. \ref{fig:8_semiclassical}, we plot the trajectories of $A(t)$ of single-tone and multi-tone readout for different time $t$ up to \qty{2.2}{\us}, and plot the integrated amplitude $\overline{A}(T)$ for different readout length $T$ up to \qty{2.2}{\us}. To give better intuition about our readout, we use experimental value of $Q_5$ (see Fig. \ref{fig:3_RFS} and Table. \ref{tab:device_params}, with $\kappa/2\pi=\qty{550}{\kHz}$, $\Omega_1/2\pi=\qty{100}{\MHz}$, $\Omega_2/2\pi=\qty{100}{\MHz}$, $\Omega_3/2\pi=\qty{75}{\MHz}$ and $\phi_d=0$.

In single-tone readout with $\omega_m=\omega_d$ (Fig. \ref{fig:8_semiclassical}(a-c)), the amplitude $A(t)$ will reach a steady state as $t\rightarrow\infty$. For multi-tone readout, the amplitude $A(t)$ will not reach a steady state no matter which rotating frame we choose. The readout result can be better understood through $\overline{A}(T)$ in Fig.\ref{fig:8_semiclassical}(g-i). The integrate amplitude $\overline{A}(T)$ reaches distinct positions in $I\!-\!Q$ plane for different transmon states. As we mentioned in the main text, only 3 to 5 trajectories are well-separated which yield high signal-noise-ratio (SNR). The other states, however, do not collapse into an exact single two-dimensional Gaussian distribution in each of these $I\!-\!Q$ planes, thus are not appropriate to be fitted using two-dimensional GMM.

\section{\label{sec:t1_analysis}$T_1$ analysis}

In general, the relaxation rate of transmon can be expressed as 
\begin{equation}
\frac{1}{T_1} = \Gamma_1 = \Gamma_{1, \text{QP}} + \Gamma_{1, \text{Purcell}} + \Gamma_{1, \text{diel}} + \Gamma_{1, \text{others}}.
\label{eq:rlx_rate_sum}
\end{equation}
For fixed frequency transmons, the relaxation can be induced by quasiparticles(QP), Purcell decay, dielectric loss. Some examples of other decay sources \cite{koch2007ChargeinsensitiveQubitDesign, smith2020} include the energy exchange with a strong two-level fluctuator(TLF) or the radiation to vacuum, which we find have relatively small contributions to our system. The values of the coherence measurement shown in Fig. \ref{fig:5_coherence} can be found in Table \ref{tab:coherence}.

\begin{table}[!t]
\caption{\label{tab:coherence} Measured coherence time of $Q_5$.}
\begin{ruledtabular}
\begin{tabular}{cccccccccccc}
Transition & $T_1$ (\unit{\us}) & $T_{2R}$ (\unit{\us}) & $T_{2E}$ (\unit{\us}) \\ 
\hline
$\ket{1} \leftrightarrow \ket{0}$ & 64(15) & 85(31) & 93(27) \\
$\ket{2} \leftrightarrow \ket{1}$ & 34(8) & 51(19) & 53(14) \\
$\ket{3} \leftrightarrow \ket{2}$ & 24(5) & 44(12) & 45(10) \\
$\ket{4} \leftrightarrow \ket{3}$ & 21(4) & 39(11) & 39(8) \\
$\ket{5} \leftrightarrow \ket{4}$ & 17(3) & 27(8) & 32(7) \\
$\ket{6} \leftrightarrow \ket{5}$ & 14(3) & 25(7) & 26(6) \\
$\ket{7} \leftrightarrow \ket{6}$ & 13(3) & 22(8) & 24(6) \\
$\ket{8} \leftrightarrow \ket{7}$ & 14(3) & 21(7) & 24(6) \\
$\ket{9} \leftrightarrow \ket{8}$ & 13(2) & 16(5) & 22(5) \\
\end{tabular}
\end{ruledtabular}
\end{table}

\subsection{\label{sec:QP}Relaxation induced by quasiparticles}

For single-junction transmons, the relaxation rate associated with the quasiparticles(QP) can be calculated by \cite{catelani2011}
\begin{equation}
\Gamma_{1, \text{QP}}^{i \rightarrow j} = | \bra{j} \sin\frac{\hat{\phi}}{2} \ket{i} |^2 \times S_{\text{QP}}(f_{ij}),
\label{eq:QP_fermi_golden_rule}
\end{equation}
and the noise spectrum density $S_{\text{QP}}(f_{ij})$ is given by
\begin{equation}
S_{\text{QP}}(f_{ij}) = x_{\text{QP}}\frac{8E_J}{\pi\hbar}\sqrt{\frac{2\Delta}{hf_{ij}}},
\label{eq:QP_spectrum_density}
\end{equation}
where $x_{\text{QP}}$ is the QP density normalized to Cooper-pair density, and $\Delta$ is the superconducting gap for thin film aluminum. For single photon decay processes, the matrix elements of $\sin\frac{\hat{\phi}}{2}$ can be approximated by
\begin{equation}
| \bra{i-1} \sin\frac{\hat{\phi}}{2} \ket{i} |^2 = \frac{i \times E_C}{hf_{01}}.
\label{eq:QP_matrix_element}
\end{equation}
To the lowest order of approximation, these matrix elements follow linear growth, which imply stronger relaxations for higher levels. Using typical parameters $\Delta \sim \qty{200}{\micro\electronvolt}$ \cite{yamamoto2006, court2008, pan2022, connolly2024}, $x_{\text{QP}} \sim \num{e-8}$, this will gives us coherence limit $T_{1, \text{QP}}^1 \sim \qty{2.2}{\ms}$ and $T_{1, \text{QP}}^9 \sim \qty{0.22}{\ms}$.

\subsection{\label{sec:Purcell}Purcell decay}

The transmon $Q_5$ does not have dedicated drive line. Here we consider the Purcell decay of a transmon through coupling to its readout resonator. The decay rate is given by \cite{koch2007ChargeinsensitiveQubitDesign}
\begin{equation}
\Gamma_{1, \text{Purcell}}^{i \rightarrow j} = 2\pi\kappa \frac{g^2 |\bra{j} \hat{n} \ket{i}|^2}{(f_i - f_j - f_r)^2}.
\label{eq:Purcell_rate}
\end{equation}
For single photon transitions, although the matrix elements of the charge operator grow for higher levels, the effective detunings $|f_{i-1, i} - f_r|$ also get larger. As the result, the Purcell decay rate may not always be enhanced. In our system, the coherence limits of Purcell decay are $T_{1, \text{Purcell}}^1 \sim \qty{271}{\us}$, $T_{1, \text{Purcell}}^7 \sim \qty{95}{\us}$, $T_{1, \text{Purcell}}^9 \sim \qty{101}{\us}$. We note that Eq. (\ref{eq:Purcell_rate}) may give an inaccurate estimation of the relaxation rate. A more careful calculation needs to consider the multi-mode nature of the transmission line resonator, and the frequency dependence of $\kappa(f)$ \cite{blaisCircuitQuantumElectrodynamics2021}.

\subsection{\label{sec:Dielectric_loss}Dielectric loss}

The relaxation through dielectric loss can be expressed by \cite{pop2014}
\begin{equation}
\Gamma_{1, \text{diel}}^{i \rightarrow j} = \frac{1}{\hbar^2} \Phi_0^2 | \bra{j} \hat{\phi} \ket{i} |^2 \times S_{\text{diel}}(f_{ij}),
\label{eq:diel_fermi_golden_rule}
\end{equation}
where $\Phi_0 = h/2e$ is the flux quantum. The spectrum density is given by
\begin{equation}
\begin{split}
S_{\text{diel}}(f_{ij}) & = \text{Re}\left[ \text{Y}(f_{ij}) \right] hf_{ij} \left[1 + \text{coth}(\frac{hf_{ij}}{2k_BT}) \right] \\
  & = \frac{2\pi f_{ij} C}{Q_{\text{diel}}(f_{ij})} hf_{ij} \left[1 + \text{coth}(\frac{hf_{ij}}{2k_BT}) \right],
\label{eq:diel_spectrum_density}
\end{split}
\end{equation}
We can relate the phase operator with the charge operator by $h f_{kj} \bra{j} \hat{\phi} \ket{k} = \text{i} 8 E_C \bra{j} \hat{n} \ket{k}$, and the decay rate can be expressed as
\begin{equation}
\Gamma_{1, \text{diel}}^{i \rightarrow j} = \frac{ 8 E_C |\bra{j} \hat{n} \ket{i}|^2 }{\hbar Q_{\text{diel}}(f_{ij})} \left[1 + \text{coth}(\frac{hf_{ij}}{2k_BT}) \right],
\label{eq:diel_rate}
\end{equation}
where we assume a phenomenological model \cite{smith2020} for the dielectric quality factor $Q_{\text{diel}}$ dependence on transition frequencies $f_{ij}$:
\begin{equation}
Q_{\text{diel}}(f_{ij}) = Q_{\text{diel}, 0} \left(  \frac{6\, \text{GHz}}{f_{ij}} \right) ^ \epsilon.
\label{eq:diel_quality_factor}
\end{equation}
With weak frequency dependence, $\Gamma_{1, \text{diel}}^{i \rightarrow j}$ usually grows up for higher energy levels. Using the typical parameters $Q_{\text{diel}, 0} \sim \num{3d6}$, $\epsilon \sim 0.7$ \cite{pop2014, smith2020, groszkowski2021}, we find the coherence limits are $T_{1, \text{diel}}^1 \sim \qty{110}{\us}$, $T_{1, \text{diel}}^9 \sim \qty{18}{\us}$.

\subsection{\label{sec:scaling}Scaling of relaxation rate}

\begin{figure}[!b]
\includegraphics{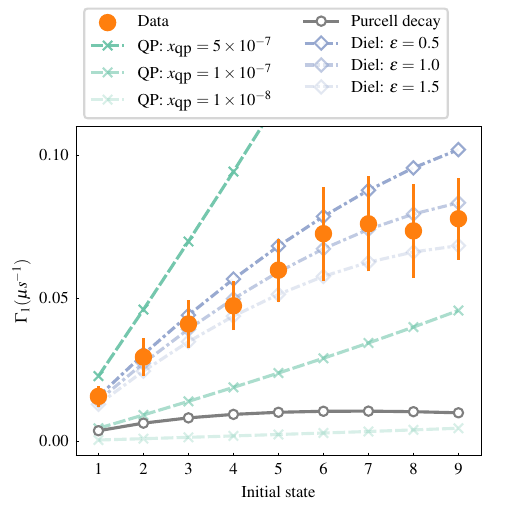}
\caption{
\label{fig:9_coherence} 
Scaling of relaxation rates for QP loss, Purcell decay and dielectric loss. The values are calculated using the equations given in Appendix \ref{sec:t1_analysis} and parameters of $Q_5$. The QP loss are evaluated for different QP density $x_{\text{QP}}$ and $\Delta = \qty{200}{\micro\electronvolt}$. The dielectric loss are evaluated for different $\epsilon$ and $Q_{\text{diel}, 0}=\num{1.8d6}$.
}
\end{figure}

Using the equations above, we can estimate how different noise sources affect the relaxation rates of higher energy levels. In Fig. \ref{fig:9_coherence}, we compare three sources where we choose different values of $x_{\text{QP}}$ for the QP loss and different values of $\epsilon$ for the dielectric loss with $Q_{\text{diel}, 0}$ fixed at $\num{1.8d6}$. We find that neither Purcell decay nor QP loss can explain our measured $T_1$ results alone. The QP loss grows up fast for higher levels, whereas the experimental results show close values when initializing transmon at $\ket{6}$ to $\ket{9}$. The dielectric loss with phenomenological parameters $\epsilon$ can better capture the scaling trend of our relaxation rates here.

To determine the parameters in Eq. (\ref{eq:diel_quality_factor}) in our system, we also treat them as fitting parameters and fit them to experimental results where we also include the QP induced relaxation with $x_{\text{QP}}=\num{e-8}$ and Purcell decay. We find $Q_{\text{diel}, 0}=\num{2.2d6}$, $\epsilon=1.2$ in our system, and the combined $\Gamma_1$ is shown in Fig.\ref{fig:5_coherence}(a). We note the result here depends on an accurate measurement of QP density $x_{\text{QP}}$, which we leave for future work.


\bibliography{HighEjEc}

\end{document}